\documentclass[aps,prd,twocolumn,groupedaddress,longbibliography]{revtex4-2}
\usepackage{amsmath}
\usepackage{bm}
\usepackage{graphicx}
\usepackage{placeins}
\usepackage{xcolor}
\usepackage[normalem]{ulem}
\usepackage{cancel}
\allowdisplaybreaks
\usepackage[hidelinks]{hyperref}
\begin{document}
	
	\title{Bardeen Black Holes in Dark Matter Halos: Effects of Inner Cutoffs and Radial Pressure on Orbital Dynamics and Inspiral Dephasing}
	
	\author{Yang Deng}
	\author{Jia-Zhou Liu}
	\author{Wen-Di Guo}
	
	\affiliation{
		Lanzhou Center for Theoretical Physics, Key Laboratory of Theoretical Physics of Gansu Province, Key Laboratory of Quantum Theory and Applications of MoE, Gansu Provincial Research Center for Basic Disciplines of Quantum Physics, Lanzhou University, Lanzhou 730000, China
	}
	\affiliation{
		Institute of Theoretical Physics \& Research Center of Gravitation, School of Physical Science and Technology, Lanzhou University, Lanzhou 730000, China
	}

\title{Radial Stress and the Innermost Stable Circular Orbit of a
	Bardeen Black Hole in a Dark Matter Halo: A First-Order Response
	Criterion}

\begin{abstract}
	A dark matter density profile alone does not determine the spacetime
	around a black hole; the radial stress must be prescribed separately
	and governs the strong-field orbital response. We study a Bardeen
	black hole in a Hernquist halo through three spherical models sharing
	the same density, mass function, and cutoff but differing in radial
	stress: a halo with $p_r^{\mathrm{DM}}=-\rho_{\mathrm{DM}}$, its
	truncated form, and a truncated Einstein cluster with
	$p_r^{\mathrm{DM}}=0$. Treating the halo as a small perturbation, we derive a
	first-order criterion for the leading innermost stable circular orbit
	shift, $C_\lambda=C_\rho+\lambda C_p$. The two truncated
	closures shift this orbit in opposite directions for the Hernquist
	profile, and unexpanded calculations for five density profiles
	confirm the predicted signs. Checks using a Hayward background and a
	published Dehnen-halo result show that the criterion is not restricted
	to the Bardeen--Hernquist system. A continuous stress interpolation identifies a critical closure at which the leading shift
	vanishes. For a representative four-year extreme-mass-ratio inspiral, changing the radial
	stress at fixed density and cutoff produces a phase difference of several radians in a
	leading-order adiabatic treatment.
\end{abstract}
	
	\maketitle
	
	\section{Introduction}
	
	Astrophysical and cosmological observations provide compelling evidence for dark matter (DM), although its microscopic nature remains unknown and different microscopic scenarios have been considered~\cite{Bertone:2004pz,Bullock:2017xww,Iocco2011,SpergelSteinhardt2000}. Strong-gravity environments, particularly the vicinity of black holes (BHs), may offer a means of probing its gravitational effects~\cite{Gondolo1999,Sadeghian2013,Jusufi:2019nrn,Bambi:2017,Psaltis:2002,BinNun:2010,Bozza2002,VirbhadraEllis2000,ClaudelVirbhadraEllis2001,Perlick2004}. In galactic nuclei, a DM halo can modify the propagation of gravitational waves (GWs) and potentially leave signatures accessible to next-generation detectors~\cite{Barausse:2014tra,Cardoso:2019rvt,Gondolo1999,Eda2015}. These prospects have motivated the construction of black-hole spacetimes immersed in DM halos~\cite{Jha2025}.
	
	A central input to such models is the DM density profile near the black hole. Gondolo and Silk studied the response of a halo to a supermassive black hole and found a DM spike with an inner depleted region~\cite{Gondolo1999,Xiong2025,Bertone:2005hw,MerrittMilos2002,UllioZhaoKamionkowski2001,Merritt2004,BertoneSiglSilk2002,GnedinPrimack2004}. For a Schwarzschild black hole, relativistic analyses locate the capture cutoff at $r=4M=2r_s$, where $r_s=2M$ is the Schwarzschild radius~\cite{Sadeghian2013,Speeney2022}. A physically credible halo model should therefore specify how the density is truncated in the strong-field region.
	
	A density profile alone does not determine the metric through the Einstein equations; a specification of the energy-momentum tensor is also required. Common spherical halo families include the Navarro--Frenk--White, Dehnen, Jaffe, and Zhao profiles~\cite{Navarro1997,Dehnen1993,Jaffe1983,Zhao1996}. A common halo closure condition is $p_r^{\rm DM}=-\rho_{\rm DM}$; together with $p_r^{\rm Bar}=-\rho_{\rm Bar}$ for the Bardeen sector, it gives the total relation $p_r=-\rho$, which, for a static spherical geometry with asymptotic normalization, implies $f(r)=g(r)$~\cite{Alloqulov:2025ucf,Jha2024,Al-Badawi:2024asn,Liang:2025vux,Zhao2026}. Cardoso \textit{et al.}\ instead used an Einstein cluster, in which the DM radial pressure vanishes and tangential stress supports circular particle motion~\cite{Cardoso2022,Lake2004,FaberVisser2006,Maeda2025}. Their analytical Hernquist model adopts an inner cutoff at $r_s$~\cite{Hernquist:1990be,Cardoso2022}. Although it has attracted considerable attention~\cite{Liu2026,Shen2024,Figueiredo2023,Pezzella2025,Fonseca2026}, the physical location of this cutoff requires further justification. Shen \textit{et al.}\ introduced a factor $(1-2r_s/r)^n$ and constructed Einstein cluster models with a depleted inner halo~\cite{Shen2024}. 	
	
	Regular black holes have been widely studied as possible resolutions of classical spacetime singularities~\cite{Ansoldi:2008jw,Bronnikov:2022ofk,Simpson:2018tsi}. Among them, the Bardeen black hole, as the first nonsingular black hole solution, was originally proposed by Bardeen~\cite{Bardeen:1968} and was later interpreted as a magnetic monopole solution of Einstein gravity coupled to nonlinear electrodynamics by Ay\'{o}n-Beato and Garc\'{i}a~\cite{Ayon-Beato:2000mjt,Toshmatov:2018}. Its physical properties and observational signatures have therefore received considerable attention~\cite{Fernando:2012yw,Pradhan:2014oaa,Lutfuoglu:2025mqa,Vertogradov:2025wgg,Abdujabbarov:2017}. More recently, regular BH solutions, including the Bardeen solution, have also been constructed within the framework of quasitopological gravity~\cite{Frolov:2024hhe,Aguayo:2025xfi}. In the present work, however, we adopt the original construction based on nonlinear electrodynamics.

	Motivated by the possibility of detecting DM signatures in strong
	gravitational fields, several studies have investigated Bardeen black
	holes surrounded by DM halos
	\cite{Narzilloev:2020qtd,Sharif:2021sow,Zhang:2020mxi,Sun2024}.
	Most existing constructions model the halo as a perfect fluid or impose
	the simplifying halo condition $p_r^{\rm DM}=-\rho_{\rm DM}$. However, even for the same
	density profile, different prescriptions for the energy-momentum tensor
	can lead to distinct spacetime geometries and observable signatures
	\cite{Fauzi:2025yse}. This issue is particularly relevant to
	extreme-mass-ratio inspirals (EMRIs), for which small environmental corrections
	to the orbital evolution may accumulate into measurable
	gravitational-wave phase differences~\cite{Barausse:2014tra,Speeney2022}.
	The accumulated dephasing of EMRI waveforms was first proposed as a
	probe of additional scalar fields by Maselli \textit{et al.}~\cite{Maselli:2020scalar}.
	The same strategy was subsequently applied and extended to dark matter
	environments by Zhao and Gong~\cite{Zhao2026}, Fu \textit{et al.}~\cite{FuEtAl2026},
	and Fier \textit{et al.}~\cite{Fier:2026yjc}. These studies showed that
	radial pressure, relativistic halo modeling, and black hole spin can
	modify the orbital evolution and accumulated dephasing. These results
	highlight the need to specify the halo
	stress tensor and the resulting spacetime geometry consistently when
	predicting strong-field and gravitational-wave observables.
	
	We therefore investigate how the choice of the DM energy-momentum
	tensor affects the geometry and observable properties of a Bardeen
	black hole surrounded by a Hernquist halo. We construct three
	models: (i) an untruncated Hernquist halo satisfying
	$p_r^{\rm DM}=-\rho_{\rm DM}$, denoted Model~1; (ii) a truncated
	halo with the same radial-stress condition, denoted Model~2; and
	(iii) a truncated Einstein cluster halo satisfying
	$p_r^{\rm DM}=0$ and supported by a self-consistent tangential
	pressure, denoted Model~3. In Models~2 and 3, the cutoff radius is
	set phenomenologically by the marginally bound circular orbit of the
	corresponding isolated Bardeen geometry. The cutoff is therefore
	determined by the black hole mass and magnetic charge rather than
	introduced as an independent halo parameter. Since Models~2 and 3
	share the same truncated density profile but obey different
	radial-stress prescriptions, their comparison directly isolates the
	geometrical effects of the assumed DM stress tensor.
	
	For the three models, we examine the energy conditions, tangential
	pressure, temporal and radial metric functions, photon sphere, and
	innermost stable circular orbit (ISCO). The comparison demonstrates
	that a density profile alone is insufficient to determine a
	spacetime containing a black hole and dark matter: the radial-pressure prescription
	also determines the temporal metric and can qualitatively change the
	strong-field orbital response. In the dilute-halo limit, we derive
	a general response criterion for the leading ISCO shift as a functional
	of the density and radial pressure at fixed magnetic charge. Comparisons
	with unexpanded ISCO calculations for standard density profiles, with and
	without an inner cutoff, test both its accuracy and its range of validity.
	Additional tests on a Hayward background and against a published
	Dehnen-halo ISCO value demonstrate that the response construction is not
	specific to the Bardeen background. Specializing
	this criterion to the Hernquist profile shows explicitly that the opposite
	ISCO shifts of Models~2 and 3 originate from their different radial
	stresses rather than from their common truncated density profile. We
	also introduce a continuous effective-stress interpolation and
	identify a critical stress parameter, dependent on the magnetic charge,
	at which the leading ISCO correction vanishes. This cancellation does
	not imply that the spacetime is unchanged, since other observables,
	including the orbital frequency, remain sensitive to the temporal
	metric.
	
	As an application, we calculate the magnetic-charge and radial-pressure
	contributions to the phase of quasicircular EMRIs using a leading-order
	adiabatic quadrupole-flux treatment. Inspirals begin at the same
	gravitational-wave frequency and phase and are compared over four years.
	For a representative galactic halo, the magnetic charge produces a
	large accumulated dephasing with a quadratic dependence at small charge,
	while changing the radial-pressure prescription at fixed density and
	charge produces a phase difference of several radians. These
	fixed-parameter calculations connect the stress dependence of the
	orbital geometry to the inspiral phase.

	This paper is organized as follows. Section~II constructs the three
	halo models, and Sec.~III compares their stresses, energy conditions,
	photon spheres, and numerical ISCO radii. Section~\ref{sec:isco_response}
	derives the first-order ISCO response, identifies the critical pressure
	closure, and tests the criterion across density profiles and
	backgrounds. Section~\ref{sec:magnetic_charge_dephasing} presents the EMRI
	framework and the phase differences caused by magnetic charge and
	radial pressure. Section~\ref{sec:summary} summarizes the results and
	their limitations. Appendix~A collects the energy-condition expressions
	for Models~1 and 2.

	\section{Construction of Metric Models}
	
	We first introduce the Bardeen solution and then construct three models of a Bardeen black hole embedded in a dark matter halo. We adopt geometric units ($G=c=1$) throughout this work.
	
	The Bardeen solution describes a static, spherically symmetric regular black hole with line element
	\begin{equation}
		ds^{2} = - f_{\text{Bar}}(r)dt^{2} + \frac{dr^{2}}{f_{\text{Bar}}(r)} + r^{2}(d\theta^{2} + \sin^{2}\theta d\phi^{2}),
	\end{equation}
	where the metric function is
	\begin{equation}
		f_{\text{Bar}}(r) = 1 - \frac{2M r^{2}}{\left( r^{2} + g_s^{2} \right)^{3/2}}.
	\end{equation}
	Here, $M$ is the BH mass and $g_s$ is the Bardeen magnetic charge parameter. When $g_s = 0$, this metric reduces to the Schwarzschild metric. From the Einstein equations, the corresponding energy density of the Bardeen geometry is
	\begin{equation}
		\rho_{\text{Bar}}(r) = \frac{3M g_s^{2}}{4\pi\left(r^{2}+g_s^{2}\right)^{5/2}}.
	\end{equation}
	
	Although collisionless cold DM is effectively pressureless on cosmological scales, a static halo in the vicinity of a black hole requires an effective-stress structure or particle orbital motion. To isolate the effect of the inner cutoff from that of the energy-momentum tensor, we construct three models using the Hernquist density profile: the first adopts the untruncated profile with $p_r^{\rm DM}=-\rho_{\rm DM}$ ($f=g$); the second introduces a cutoff while keeping $p_r^{\rm DM}=-\rho_{\rm DM}$; and the third uses the truncated profile with the Einstein cluster closure $p_r^{\rm DM}=0$ (anisotropic stress, $f\neq g$). This setup makes the construction readily transferable to other spherical halo profiles.
	
	\subsection{Model 1: Standard $p_r^{\rm DM}=-\rho_{\rm DM}$ Approach}
	The DM is assumed to be a fluid with energy-momentum tensor of the form
	\begin{equation}
		T^{\mu}{}_{\nu}^{\rm DM}
		=
		\mathrm{diag}\bigl(-\rho_{\rm DM}(r),\ p_r^{\rm DM}(r),\ p_t^{\rm DM}(r),\ p_t^{\rm DM}(r)\bigr),
	\end{equation}
	where $p_r^{\rm DM}=-\rho_{\rm DM}$, while the DM tangential pressure $p_t^{\rm DM}(r)$ is fixed by the field equations. Together with $p_r^{\rm Bar}=-\rho_{\rm Bar}$, this condition gives the total closure $p_r=-\rho$ and implies, after asymptotic normalization, that $f_1(r)=g_1(r)$, or equivalently $g_{tt}(r)=-1/g_{rr}(r)$. The line element therefore becomes
	\begin{equation}
		ds_{1}^{2}
		=
		-f_{1}(r)\,dt^{2}
		+\frac{dr^{2}}{f_{1}(r)}
		+r^{2}\left(d\theta^{2}+\sin^{2}\theta\,d\phi^{2}\right).
	\end{equation}
	The subscript 1 labels the metric functions of Model~1.
	
	We model the dark matter density using the Hernquist profile~\cite{Hernquist:1990be}:
	\begin{equation}
		\rho_{\rm DM}(r)
		=
		\frac{M_{\rm DM}a_{0}}
		{2\pi r(a_{0}+r)^{3}},
	\end{equation}
	where $M_{\rm DM}$ denotes the total dark matter mass and $a_{0}$ is the characteristic scale radius of the halo.
	
	Several previous studies infer the metric function from a Newtonian approximation~\cite{Liang:2025vux,Al-Badawi:2024asn,Gohain:2024eer,Matos2003}. Bolokhov has emphasized the limitations of extrapolating a Newtonian construction into the strong-field regime and the need to solve the Einstein equations directly~\cite{Bolokhov2025}. We therefore determine the metric from the Einstein equations.
	
	Introducing the mass function
	\begin{equation}
		g_1(r)=1-\frac{2m_1(r)}{r},
	\end{equation}
	the field equations become~\cite{Wald:1984}
	\begin{equation}
		\frac{m'_1(r)}{r^{2}}
		=
		4\pi\rho_{1}(r),
	\end{equation}
	
	\begin{equation}
		\frac{-2f_1(r)m_1(r)+r\left[r-2m_1(r)\right]f_1'(r)}
		{r^{3}f_1(r)}
		=
		8\pi p_{r,1}(r),
	\end{equation}
	
	\begin{equation}
		\begin{aligned}
			8\pi p_{t,1}(r)
			={}&\frac{1}{4r^3f_1^2}
			\Bigg\{
			-r^2(r-2m_1)f_1'^{\,2}
			\\
			&\quad+4f_1^2(m_1-rm_1')
			\\
			&\quad
			+2rf_1
			\Big[
			-f_1'\left(m_1+r(-1+m_1')\right)
			\\
			&\qquad
			+r(r-2m_1)f_1''
			\Big]
			\Bigg\}.
		\end{aligned}
	\end{equation}
	
	Because the Bardeen and DM sectors are assumed to couple only through gravity, the total energy-momentum tensor is
	\begin{equation}
		T_{\mu\nu}
		=
		T_{\mu\nu}^{\rm Bar}
		+
		T_{\mu\nu}^{\rm DM}.
	\end{equation}
	
	The corresponding total energy density is
	
	\begin{equation}
		\begin{aligned}
			\rho_{1}(r)
			&=
			\rho_{\rm Bar}(r) + \rho_{\rm DM}(r) \\
			&=
			\frac{3Mg_{s}^{2}}
			{4\pi(r^{2}+g_{s}^{2})^{5/2}}
			+
			\frac{M_{\rm DM}a_{0}}
			{2\pi r(a_{0}+r)^{3}}.
		\end{aligned}
	\end{equation}
	
	Integrating the Hernquist profile gives $m_{\rm DM}(r)=M_{\rm DM}r^2/(a_0+r)^2$. Substituting this function into the field equations yields the tangential pressure $p_t(r)$. The metric function for Model~1 is therefore
	
	\begin{equation}
		f_{1}(r)
		=
		1
		-
		\frac{2Mr^{2}}
		{(g_{s}^{2}+r^{2})^{3/2}}
		-
		\frac{2M_{\rm DM}r}{(a_0+r)^2}.
	\end{equation}
	
	We next examine the energy conditions. For the parameter values considered below, the weak energy condition (WEC) and null energy condition (NEC) are satisfied outside the horizon. The expressions used to evaluate the dominant energy condition (DEC) and strong energy condition (SEC) are collected in Appendix~A.
	
	\begin{figure}[t]
		\centering
		
		\includegraphics[width=0.88\columnwidth]{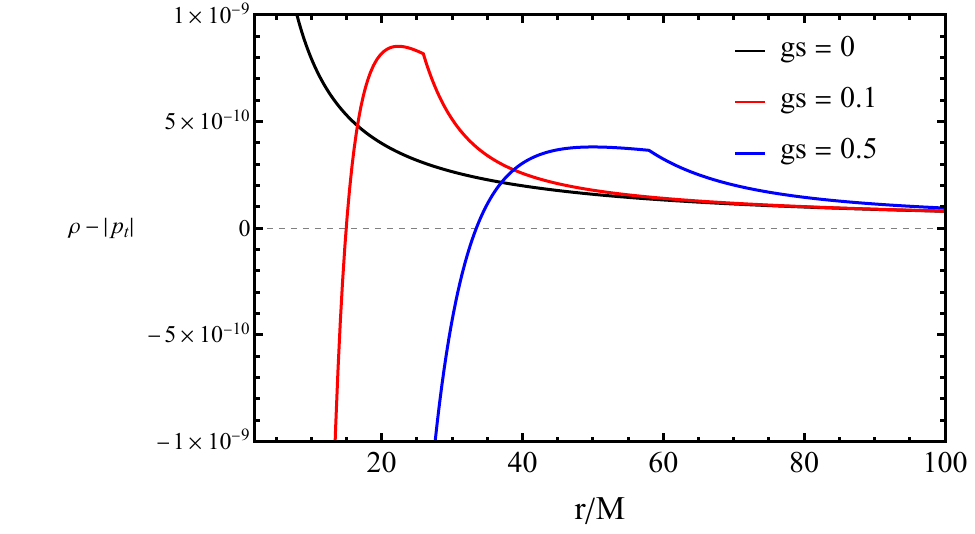}
		
		\vspace{0.3cm}
		
		\includegraphics[width=0.88\columnwidth]{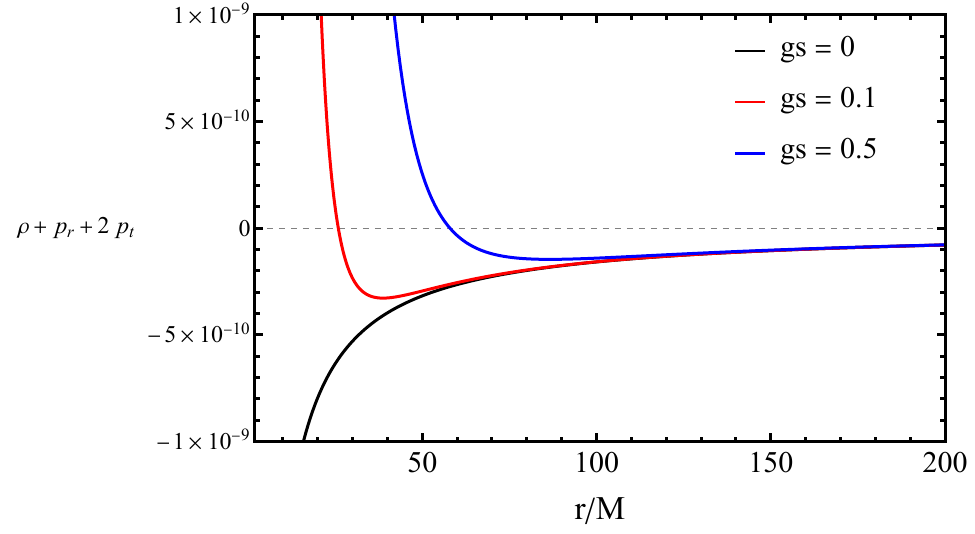}
		
		\caption{
			Radial profiles of the dominant and strong energy conditions for Model~1.
			The upper panel shows $\rho-|p_t|$, corresponding to the nontrivial DEC,
			while the lower panel shows $\rho+p_r+2p_t$, corresponding to the SEC.
			The vertical axes have dimensions of $1/M^2$; numerically we set $M=1$, so the plotted values are dimensionless. The same normalization is used for all subsequent energy-condition plots.
			The parameters are
			$M_{\rm DM}=10^{3}$,
			$a_0=10^{5}$,
			and
			$g_s=0,\;0.1,\;0.5$.
		}
		\label{fig:model1_energy}
	\end{figure}
	
	The top panel of Fig.~\ref{fig:model1_energy} shows the radial behavior of $\rho-|p_t|$. DEC is satisfied when this quantity is nonnegative and violated when it is negative. This quantity is negative only in the vicinity of the black hole, indicating a local violation of the DEC. As the radius increases, it rapidly becomes positive and asymptotically approaches zero from above. Moreover, increasing the magnetic charge shifts the zero-crossing radius outward, implying that the region in which the DEC is violated becomes more extended.
	
	The bottom panel of Fig.~\ref{fig:model1_energy} displays the quantity $\rho+p_r+2p_t$. SEC is satisfied when this quantity is nonnegative and violated when it is negative. For nonzero magnetic charge, it is positive near the black hole but changes sign at a finite radius and subsequently approaches zero from below. In contrast, for the Schwarzschild limit ($g_s=0$), it remains negative throughout the plotted range. Therefore, increasing the magnetic charge enlarges the inner region in which the SEC is satisfied, while all curves converge toward zero at large radii as the influence of both the nonlinear electromagnetic field and the Hernquist dark matter halo gradually diminishes.
	
	\subsection{Model 2: Modified $p_r^{\rm DM}=-\rho_{\rm DM}$ with Cutoff Radius}
	
	In Model~1, the Hernquist profile contributes to the mass function outside the event horizon, including the strong-field region. To model the expected depletion of DM particles, we impose an inner cutoff outside the outer horizon. Cardoso \textit{et al.}\ used the horizon as this cutoff, but particles near the horizon are inevitably captured by the black hole~\cite{Cardoso2022}. We instead adopt the marginally bound circular orbit as a phenomenological cutoff criterion. This recovers $r_t=4M$ for a Schwarzschild black hole~\cite{Sadeghian2013} and can be generalized to the Bardeen geometry.
	
	Specifically, we set the DM density to zero at and below the marginally bound radius $r_t$. The marginally bound radius $r_t$ is obtained from the circular-orbit conditions for a unit-mass particle with $\mathcal{E}=1$ in the Bardeen spacetime.
	
	The conserved quantities of a unit-mass test particle are the energy $\mathcal{E} = f_{\text{Bar}}(r)\dot{t}$ and the angular momentum $L = r^{2}\dot{\phi}$. From $g_{\mu\nu}\dot{x}^{\mu}\dot{x}^{\nu} = -1$, the radial equation is obtained:
	\begin{equation}
		\dot{r}^{2} = \mathcal{E}^{2} - f_{\text{Bar}}(r)\left( 1+\frac{L^{2}}{r^{2}} \right) \equiv \mathcal{E}^{2} - V_{\text{eff}}(r),
	\end{equation}
	\begin{equation}
		V_{\text{eff}}(r) = f_{\text{Bar}}(r)\left( 1+\frac{L^{2}}{r^{2}} \right).
	\end{equation}
	Circular orbits require $V_{\text{eff}}'(r)=0$ and $V_{\text{eff}}(r)=\mathcal{E}^{2}$. Differentiating the effective potential gives
	\begin{equation}
		V_{\text{eff}}'(r) = f_{\text{Bar}}'(r)\left( 1+\frac{L^{2}}{r^{2}} \right) + f_{\text{Bar}}(r)\left( -\frac{2L^{2}}{r^{3}} \right) = 0.
	\end{equation}
	Thus,
	\begin{equation}
		f_{\text{Bar}}'(r)r^{3}\left( 1+\frac{L^{2}}{r^{2}} \right) - 2L^{2}f_{\text{Bar}}(r) = 0.
	\end{equation}
	After rearrangement, the expression for $L^{2}$ is
	\begin{equation}
		L^{2} = \frac{f_{\text{Bar}}'(r)r^{3}}{2f_{\text{Bar}}(r) - f_{\text{Bar}}'(r)r}.
	\end{equation}
	For the Bardeen metric, differentiating $f_{\text{Bar}}(r)$ gives
	\begin{equation}
		f_{\text{Bar}}'(r)   = \frac{2M r(r^{2} - 2g_s^{2})}{\left( r^{2} + g_s^{2} \right)^{5/2}}.
	\end{equation}
	The circular-orbit condition $V_{\text{eff}}(r)=\mathcal{E}^{2}$ gives
	\begin{equation}
		\mathcal{E}^{2} = 1= \frac{2f_{\text{Bar}}(r)^{2}}{2f_{\text{Bar}}(r) - f_{\text{Bar}}'(r)r}.
	\end{equation}
	
	For the Bardeen metric, the corresponding nonlinear algebraic equation is solved numerically for each value of $g_s/M$. When more than one positive root occurs, the physically relevant choice must be the outer root lying outside the outer horizon.
	
\begin{figure}[htbp]
		\centering
		\includegraphics[width=0.4\textwidth]{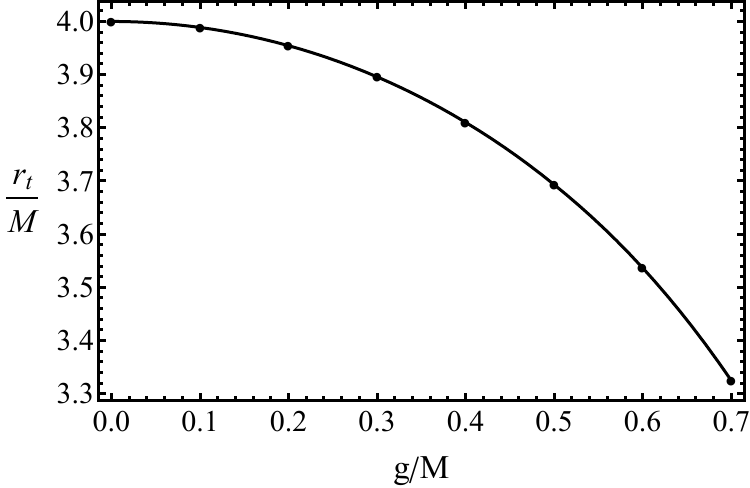}
		\caption{The cutoff radius \(r_t\) as a function of the magnetic charge parameter \(g_s\). The curve is obtained by numerically solving the marginally bound orbit condition for the Bardeen black hole.}
		\label{fig:rt}
	\end{figure}

	For the outer branch displayed in Fig.~\ref{fig:rt}, $r_t$ decreases from its Schwarzschild value as $g_s$ increases. This trend reflects the modification of the Bardeen gravitational potential by the magnetic charge parameter. 
	
	We prescribe the DM density to vanish in the inner region. Following Shen \textit{et al.}, we multiply the Hernquist profile by $(1-r_t/r)$ outside the cutoff, so that the density vanishes continuously at $r=r_t$:
	\begin{equation}
		\rho_{\text{DM},2}(r) =
		\begin{cases}
			0, & r < r_t, \\[4pt]
			\dfrac{M_{\text{DM}} a_0 \left( 1 - \dfrac{r_t}{r} \right)}{2\pi r (a_0 + r)^3}, & r \geq r_t.
		\end{cases}
		.
	\end{equation}
	The integrated mass of the truncated halo is
	$M_{\rm halo}=M_{\rm DM}a_0/(a_0+r_t)$. For the parameter ranges
	considered here, $r_t/a_0\ll1$, so that $M_{\rm halo}$ differs
	negligibly from $M_{\rm DM}$.
	The total energy density for Model~2 is
	\begin{equation}
		\rho_{2}(r) =
		\begin{cases}
			\dfrac{3M g_s^{2}}{4\pi\left( r^2 + g_s^{2}\right)^{5/2}}, & r < r_t, \\[4pt]
			\dfrac{M_{\text{DM}} a_0 \left( 1 - \dfrac{r_t}{r} \right)}{2\pi r (a_0 + r)^3} + \dfrac{3M g_s^{2}}{4\pi\left(r^2+g_s^{2}\right)^{5/2}}, & r \geq r_t.
		\end{cases}
		.
	\end{equation}
	Integrating the field equations with the inner boundary condition gives the piecewise metric function
	\begin{equation}
		f_{2}(r)=
		\begin{cases}
			f_{\rm Bar}(r), & r < r_t, \\[4pt]
			f_{\rm Bar}(r)-\dfrac{2a_0 M_{\rm DM}(r-r_t)^2}{r(a_0+r)^2(a_0+r_t)} , & r \ge r_t.
		\end{cases}
		,
	\end{equation}
	where the result is obtained by following the same integration procedure as in Model~1, with the truncated density profile given above.
	
\begin{figure}[htbp]
		\centering
		\includegraphics[width=0.45\textwidth]{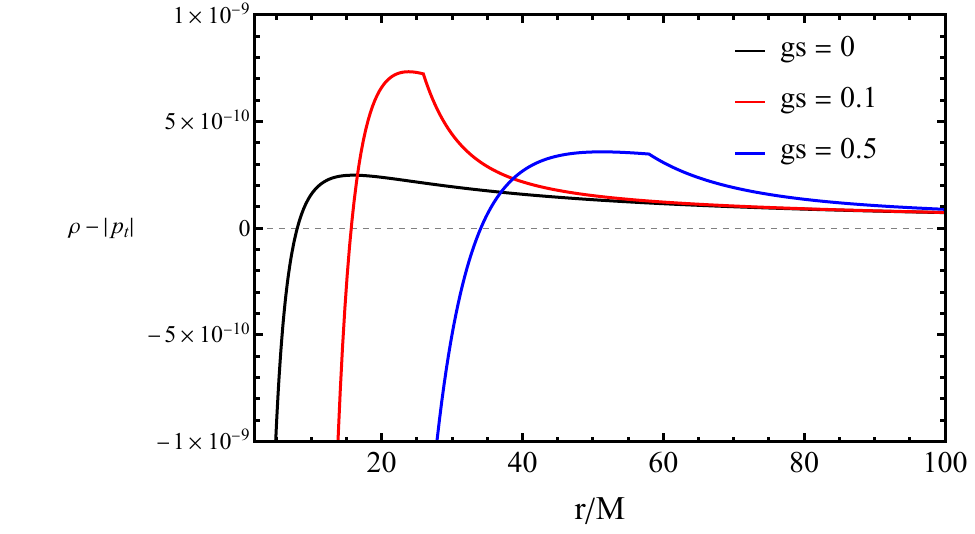}
		\includegraphics[width=0.45\textwidth]{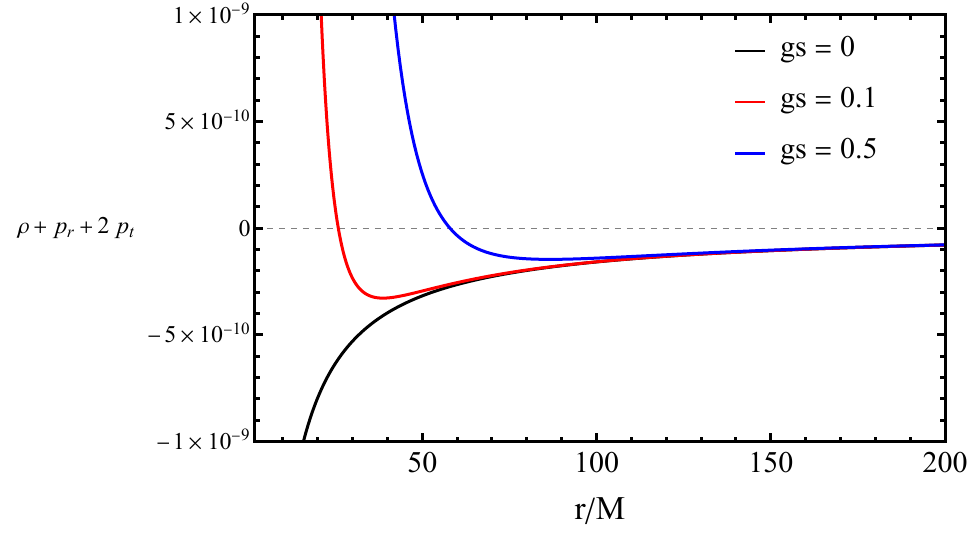}
		\caption{
			Radial profiles of the dominant and strong energy conditions for Model 2.
			The upper panel shows $\rho-|p_t|$, corresponding to the nontrivial DEC,
			while the lower panel shows $\rho+p_r+2p_t$, corresponding to the nontrivial SEC.
			The parameters are
			$M_{\rm DM}=10^{3}$,
			$a_0=10^{5}$,
			and
			$g_s=0,\;0.1,\;0.5$.
		}
		\label{fig:model2_energy}
	\end{figure}

For the representative parameters in Fig.~\ref{fig:model2_energy}, the WEC and NEC are satisfied outside the cutoff for \(g_s=0.1\) and \(0.5\). In the Schwarzschild limit \(g_s=0\), however, \(\rho+p_t<0\) in a narrow interval immediately outside \(r_t\), resulting in localized violations of both the WEC and NEC. The SEC and DEC are also not satisfied throughout the exterior region: the DEC is violated close to \(r_t\), while the SEC is negative over an intermediate radial interval before recovering asymptotically. The cutoff removes the prescribed inner DM component, but it does not by itself ensure all of the energy conditions. The analytic expressions used for these diagnostics are provided in Appendix~A.
	
	\subsection{Model 3: Einstein Cluster Assumption}
	
	The first two models impose $p_r^{\rm DM}=-\rho_{\rm DM}$, which is a convenient ansatz for the DM energy-momentum tensor but not the only possible one. We next adopt the Einstein cluster construction of Einstein and of Cardoso \textit{et al.}~\cite{Einstein:1939ms,Cardoso2022}. This is a collisionless steady-state configuration: individual DM particles move on circular orbits, while their ensemble has a time-independent density and energy-momentum tensor. The DM radial pressure therefore vanishes and the tangential stress is associated with circular particle motion. Its energy-momentum tensor is
	\begin{equation}
		T^{\mu}{}_{\nu}^{\rm DM}=\mathrm{diag}\bigl(-\rho_{\rm DM},0,p_t^{\rm DM},p_t^{\rm DM}\bigr).
	\end{equation}
	The spacetime metric for Model 3 is written as
	\begin{equation}
		ds_{3}^{2} = - f_{3}(r)dt^{2} + \frac{1}{g_3(r)} dr^{2} + r^{2}(d\theta^{2} + \sin^{2}\theta d\phi^{2}).
	\end{equation}
	Here, $g_3(r)=1-\frac{2m_3(r)}{r}$.
	
	The tangential pressure follows from the Einstein equations and energy-momentum conservation. The relevant relations are
	\begin{equation}
		\frac{m'_3(r)}{r^{2}}
		=
		4\pi\rho_3(r),
	\end{equation}
	
	\begin{equation}
		\frac{f'_3}{f_3}=\frac{-8\pi r^3\rho_{Bar}(r)+2m_3(r)}{r(r-2m_3(r))}
	\end{equation}
	
	\begin{equation}
		p_{t}(r)
		=
		p_{r}^{\rm Bar}(r)
		+\frac{r}{2}\frac{d p_{r}^{\rm Bar}(r)}{dr}
		+\frac{r\,\rho_{\rm DM}(r)}{4}
		\frac{f'_3(r)}{f_3(r)}.
	\end{equation}
	
	We use the same cutoff density profile in the Einstein cluster construction. The total energy density is
	\begin{equation}
		\rho_{3}(r) = 
		\begin{cases}
			\dfrac{3M g_s^{2}}{4\pi\left(r^2+g_s^2\right)^{5/2}}, & r < r_t, \\[10pt]
			\dfrac{3M g_s^{2}}{4\pi\left(r^2+g_s^2\right)^{5/2}} + \dfrac{M_{\text{DM}} a_0 \left( 1 - \dfrac{r_t}{r} \right)}{2\pi r (a_0 + r)^3}, & r \geq r_t.
		\end{cases}
	\end{equation}
	
	After substituting the total energy density $\rho_3(r)$ into the
	field equations, we solve numerically for $f_3(r)$ with
	$f_3(\infty)=1$. The interior and exterior solutions are matched
	continuously at the cutoff radius $r_t$, following the procedure of
	Ref.~\cite{Shen2024}. The resulting matching factor represents the
	constant gravitational redshift of the inner Bardeen region induced
	by the exterior halo.
	
	\begin{figure}[htbp]
		\centering
		\includegraphics[width=0.45\textwidth]{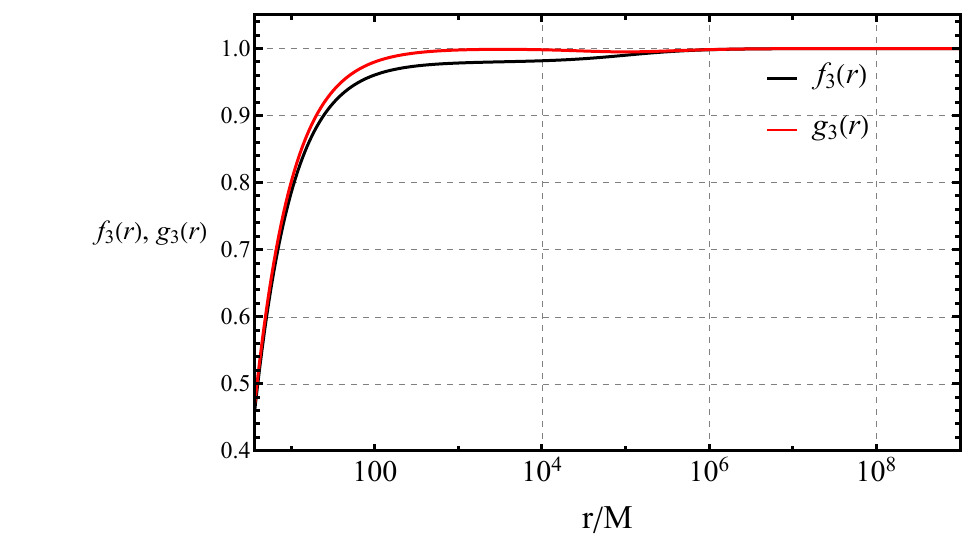}
		\caption{
			Radial dependence of the metric functions $f_3(r)$ and $g_3(r)$ for Model 3.
			The parameters are
			$M_{\rm DM}=10^{3}$,
			$a_0=10^{5}$,
			and
			$g_s=0.5$.
			The comparison shows the deviation between the temporal and radial metric functions introduced by the Einstein cluster description.
		}
		\label{fig:model3_metric}
	\end{figure}
	
	\begin{figure}[htbp]
		\centering
		\includegraphics[width=0.45\textwidth]{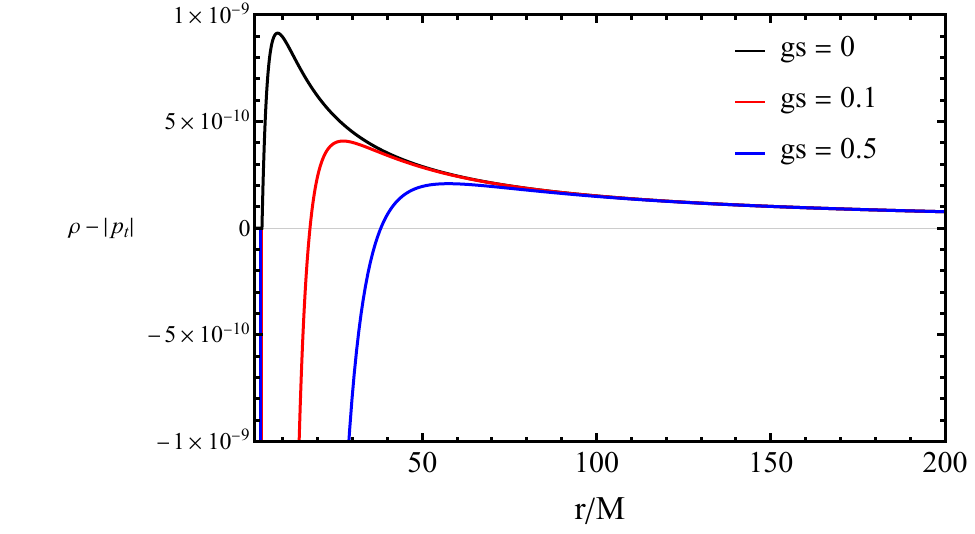}
		\caption{
			Radial profile of the DEC for Model 3.
			The quantity $\rho-|p_t|$ is shown for different magnetic charge parameters.
			The DEC violation is confined to a narrow region near the cutoff radius and rapidly approaches positive values at larger radii.
			The parameters are
			$M_{\rm DM}=10^{3}$ and $a_0=10^{5}$.
		}
		\label{fig:model3_dec}
	\end{figure}
	
	Figure~\ref{fig:model3_metric} shows the radial profiles of the metric functions $f_3(r)$ and $g(r)$ for Model~3. In the inner region $r<r_t$, where the dark matter density is set to
	zero, the radial metric coincides with that of the Bardeen solution, while the temporal metric
	differs only by the constant redshift factor fixed through matching
	to the exterior halo. Outside the cutoff radius, $f_3(r)$ and $g(r)$ deviate from each other, reflecting the anisotropic stress characteristic of the Einstein cluster construction. At large radii, both functions asymptotically approach unity, consistent with asymptotic flatness.
	
	Figure~\ref{fig:model3_dec} shows the Model 3 energy-condition diagnostics. In the numerical range displayed, the WEC, NEC, and SEC are satisfied, whereas the DEC is violated in a finite interval immediately outside \(r_t\) and is restored at larger radii. Both the magnitude and the radial extent of this violation increase with \(g_s\) over the parameter range considered here. This behavior originates from the Bardeen nonlinear electrodynamic sector, while the positive tangential stress supplied by the Einstein cluster halo gradually restores the total DEC at larger radii.

	\section{Comparative Analysis of the Three Models}
	
	\subsection{Comparison of Tangential Pressures}

\begin{figure}[htbp]
		\centering
		\includegraphics[width=0.72\columnwidth]{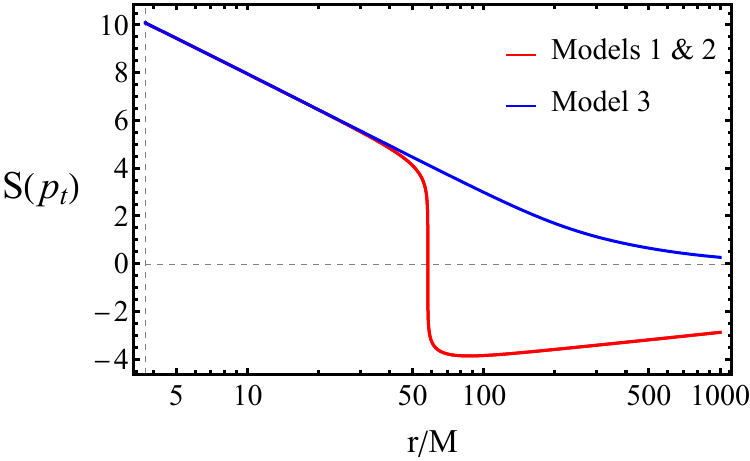}
		\caption{
			Comparison of the tangential-pressure profiles of the three models for
			$M=1$, $M_{\rm DM}=10^{3}$, $a_0=10^{5}$, and $g_s=0.5$.
			Models~1 and 2 are shown in red and are nearly coincident, while the
			Einstein cluster Model~3 is shown in blue. To display both signs over a
			wide dynamic range, the vertical coordinate is
			$S(p_t)=\operatorname{sgn}(p_t)\log_{10}[1+\frac{|p_t|}{10^{-14}M^{-2}}]$.
			The underlying tangential pressure $p_t$ is given in units of
			$M^{-2}$.
			The dashed horizontal line denotes $p_t=0$: Models~1 and 2 become
			negative outside their zero-crossing, whereas Model~3 remains positive
			in the displayed interval.
		}
		\label{fig:tangential_pressure}
	\end{figure}

	The tangential-pressure profiles provide a direct comparison of the energy-momentum tensors employed in the three models. For Models~1 and 2, the total closure is $p_r=-\rho$, whereas in Model~3 only the DM radial pressure vanishes and the Bardeen sector retains its intrinsic anisotropic stress. Figure~\ref{fig:tangential_pressure} displays the resulting profiles.

	The near overlap of Models~1 and 2 follows from their common Bardeen contribution and the large halo scale: in the inner halo region $r_t\leq r\ll a_0$, the cutoff changes the DM tangential pressure contribution by a relative correction of approximately $3r_t/a_0$. Both profiles cross zero near $r\simeq 58M$ and subsequently become negative. This is precisely the kink seen in the preceding DEC curves. For Models~1 and 2, the relevant DEC combination is $\rho-|p_t|$. At $p_t=0$, the absolute value changes between its $+p_t$ and $-p_t$ branches, so $\rho-|p_t|$ is continuous but changes slope precisely at the zero of $p_t$. The apparent DEC turning point is therefore not an additional physical radius; it directly records the reversal of the tangential stress.

	The same radius also has a direct SEC interpretation. Indeed, for the $p_r=-\rho$ closure used in Models~1 and 2, the strong energy condition combination reduces to
	\begin{equation}
		\rho+p_r+2p_t=2p_t .
	\end{equation}
	Consequently, the radial location at which the SEC curve changes sign is also exactly the zero of $p_t$. The negative branch of $p_t$ is therefore the direct origin of the outer SEC-violating region in these two models.

	By contrast, the Einstein cluster closure produces a positive tangential pressure throughout the displayed exterior region. This sign has a clear physical origin: the DM particles have no radial velocity dispersion ($p_r^{\rm DM}=0$), but move on circular orbits, whose angular momentum supplies a positive transverse stress. In the field equation expression
	\begin{equation}
		p_{t,3}=p_r^{\rm Bar}+\frac{r}{2}\frac{d p_r^{\rm Bar}}{dr}
		+\frac{r\rho_{\rm DM}}{4}\frac{f_3'}{f_3},
	\end{equation}
	the last term is this orbital-support contribution. Outside $r_t$ it is positive for the present solution ($\rho_{\rm DM}>0$ and $f_3'/f_3>0$); together with the positive Bardeen tangential contribution in this radial range, it keeps $p_{t,3}>0$. Thus, the Model~3 profile represents the stress signature of a collisionless circular-orbit cluster. This also explains why its SEC remains satisfied in the corresponding energy-condition plot. This qualitative difference is obscured on an ordinary logarithmic plot, which cannot represent negative values, and is therefore shown using the signed logarithmic coordinate in Fig.~\ref{fig:tangential_pressure}.

	\subsection{Comparison of Energy Conditions}
	
	The energy conditions provide a diagnostic of the effective matter source and of the consistency of the adopted energy-momentum tensor. In the numerical examples considered here, Models 1 and 3 satisfy the WEC and NEC outside the relevant inner boundary. Model 2 also satisfies these conditions for the nonzero magnetic charges shown, whereas in the Schwarzschild limit it exhibits localized WEC and NEC violations immediately outside the cutoff.
	
	For nonzero magnetic charge, Models 1 and 2 exhibit an extended DEC-violating interval that begins in the vicinity of the black hole and terminates at a zero-crossing radius \(r_0\), which increases with \(g_s\). This extended violation originates predominantly from the Bardeen nonlinear electrodynamic sector. In Model 1, the dark matter component does not produce an additional localized violation. Model 2, however, contains a separate violation caused by the cutoff: although the truncated density vanishes continuously at \(r_t\), its rapid growth immediately outside the cutoff produces a narrow region in which the dark matter sector violates the energy conditions. In the Schwarzschild limit, Model 1 satisfies the DEC throughout the exterior region, whereas Model 2 retains this localized violation caused by the cutoff. The SEC of both models is violated over an intermediate radial interval, with the dark matter contribution providing the corresponding negative tangential-pressure branch.
	
	Model~3 behaves differently. Its SEC is satisfied throughout the exterior
	region, while its DEC violation is confined to a narrow interval immediately
	outside $r_t$ (Fig.~\ref{fig:model3_dec}), with both its magnitude and radial
	extent increasing as $g_s$ increases. An isolated Einstein cluster halo can satisfy
	the weak, strong, and dominant energy conditions when its inner radius is
	sufficiently large~\cite{Shen2025}, and the truncated Einstein cluster
	models constructed in Ref.~\cite{Shen2024} satisfy all three conditions.
	In the present geometry containing the Bardeen black hole and DM, the localized DEC violation
	is instead inherited from the Bardeen nonlinear electrodynamic sector,
	which is known to violate the dominant energy condition in part of the
	spacetime~\cite{Lan:2022bld}. Because both the truncated dark matter density and
	its orbital-support contribution to the tangential pressure vanish at
	$r=r_t$, the Einstein cluster halo cannot compensate for the Bardeen sector
	violation immediately outside the cutoff. As the halo contribution grows
	with radius, the total DEC is restored. The cutoff therefore controls the
	radial extent of the DEC-violating region rather than its physical origin.
	
	Thus, a physically motivated density cutoff and a physically motivated
	ansatz for the energy-momentum tensor address distinct aspects of the construction. The
	cutoff improves the interpretation of the inner halo, whereas the
	energy-momentum tensor controls how the density profile sources the metric and
	therefore has a direct impact on the energy conditions. The comparison
	confirms that neither the density profile nor the cutoff radius alone
	determines the physical viability of a spacetime containing a black hole and dark matter;
	the full energy-momentum tensor must be specified.
	
	\subsection{Numerical photon-sphere and ISCO results}
	
	We now compare the photon spheres and ISCOs of the three metrics with those
	of the isolated Bardeen black hole. For a static, spherically symmetric
	metric
	\begin{equation}
		ds^2 = -f(r)\,dt^2 + \frac{dr^2}{g(r)}
		+ r^2\bigl(d\theta^2 + \sin^2\theta\,d\phi^2\bigr),
		\label{eq:metric-general}
	\end{equation}
	the photon sphere $r_{\rm ps}$ is the radius of the unstable circular null
	orbit, obtained from $d(f/r^2)/dr = 0$, i.e.
	\begin{equation}
		r f'(r) - 2 f(r) = 0 .
		\label{eq:photon-sphere}
	\end{equation}
	For a timelike test particle with conserved energy $E$ and angular momentum
	$L$~\cite{Chandrasekhar:1983}, the radial equation reads
	\begin{equation}
		\dot r^2 = g(r)\left[\frac{E^2}{f(r)} - \left(1 + \frac{L^2}{r^2}\right)\right],
		\label{eq:radial}
	\end{equation}
	so that circular orbits satisfy $V_{\rm eff}'(r) = 0$ and the ISCO further
	requires $V_{\rm eff}''(r) = 0$, where
	\begin{equation}
		V_{\rm eff}(r) = f(r)\left(1 + \frac{L^2}{r^2}\right),\qquad
		L^2 = \frac{r^3 f'(r)}{2 f(r) - r f'(r)} .
		\label{eq:isco}
	\end{equation}
	Because $g(r)$ enters Eq.~\eqref{eq:radial} only as an overall factor, it
	cancels in the circular orbit and stability conditions; hence
	$r_{\rm ps}$ and $r_{\rm ISCO}$ are determined by the temporal metric
	function $f(r)$ alone. For the Einstein cluster metric ($g_3 \neq f_3$) this
	is an important simplification: the orbital radii are insensitive to the
	radial component $g(r)$ and probe only $f(r)$.
	
	Table~\ref{tab:isco} lists the cutoff, photon sphere, and ISCO radii for a
	parameter scan in the halo mass and magnetic charge. The halo scale is fixed
	at $a_0=10^4$, while $M_{\rm DM}=10^2$, $10^3$, and $10^4$ and
	$g_s=0$, $0.3$, and $0.5$ are varied. The Model~1 photon-sphere radius
	$r_{\rm ps}^{\rm M1}$ is obtained numerically from
	$r f_1'(r)-2f_1(r)=0$, selecting the outer unstable root. For the displayed
	parameters, the pure Bardeen photon sphere lies inside the cutoff radius,
	$r_{\rm ps}^{\rm Bar}<r_t$. Hence, in the photon-sphere region,
	$f_2(r)=f_{\rm Bar}(r)$ and $f_3(r)=C f_{\rm Bar}(r)$, where
	$C$ is the constant matching factor. Since $C$ cancels from
	Eq.~(33), Models~2 and~3 reproduce the pure Bardeen
	photon-sphere radius. Model~1 has no inner cutoff, so its
	photon-sphere radius receives a small correction from the Hernquist halo.
	\begin{table*}[ht]
		\centering
		\caption{
			Cutoff, photon sphere, and ISCO radii for the scan over halo mass and magnetic charge.
			Here $a_0=10^4$, $M=1$, and all entries are given in units of $M$.
		}
		\label{tab:isco}
		\scriptsize
		\begin{ruledtabular}
			\begin{tabular}{ccccccccc}
				$M_{\rm DM}$ &
				$g_s$ &
				$r_t$ &
				$r_{\rm ps}^{\rm Bar}$ &
				$r_{\rm ISCO}^{\rm Bar}$ &
				$r_{\rm ps}^{\rm M1}$ &
				$r_{\text{ISCO}}^{(\text{Model 1})}$ &
				$r_{\text{ISCO}}^{(\text{Model 2})}$ &
				$r_{\text{ISCO}}^{(\text{Model 3})}$ \\
				\hline
				$10^2$ & 0.0 & 4.00000 & 3.00000 & 6.00000 & 3.00001 & 6.00022 & 6.00022 & 5.99988 \\
				$10^2$ & 0.3 & 3.89578 & 2.92240 & 5.85376 & 2.92241 & 5.85397 & 5.85397 & 5.85365 \\
				$10^2$ & 0.5 & 3.69306 & 2.76871 & 5.57222 & 2.76872 & 5.57240 & 5.57240 & 5.57212 \\
				$10^3$ & 0.0 & 4.00000 & 3.00000 & 6.00000 & 3.00009 & 6.00216 & 6.00216 & 5.99880 \\
				$10^3$ & 0.3 & 3.89578 & 2.92240 & 5.85376 & 2.92249 & 5.85579 & 5.85579 & 5.85262 \\
				$10^3$ & 0.5 & 3.69306 & 2.76871 & 5.57222 & 2.76880 & 5.57402 & 5.57402 & 5.57117 \\
				$10^4$ & 0.0 & 4.00000 & 3.00000 & 6.00000 & 3.00090 & 6.02189 & 6.02188 & 5.98814 \\
				$10^4$ & 0.3 & 3.89578 & 2.92240 & 5.85376 & 2.92330 & 5.87433 & 5.87430 & 5.84241 \\
				$10^4$ & 0.5 & 3.69306 & 2.76871 & 5.57222 & 2.76962 & 5.59045 & 5.59035 & 5.56175 \\
			\end{tabular}
		\end{ruledtabular}
		\normalsize
	\end{table*}

	The ISCO responds differently from the photon sphere. Relative to the
	pure Bardeen black hole, Models~1 and~2 shift the ISCO outward, whereas
	Model~3 shifts it inward. Models~2 and~3 have the same truncated density
	profile, so their opposite ISCO shifts isolate the role of the
	energy-momentum tensor: the circular-orbit correction depends not only on the density
	profile, but also on the energy-momentum tensor that sources the metric.

	Because $r_{\rm ISCO}>r_t$, the ISCO directly probes the exterior halo.
	For the present $a_0=10^4$ scan, the difference between Models~1 and~2
	remains small, but becomes resolvable at the displayed precision for larger
	$M_{\rm DM}$. It is caused by the different halo mass contributions in the
	ISCO region for the untruncated and truncated Hernquist profiles. In
	contrast, the inward Model~3 shift originates from its different
	radial-stress closure, $p_{r,3}^{\rm DM}=0$, which modifies the temporal
	metric $f_3(r)$ even though Models~2 and~3 share the same density profile
	and mass function. The associated tangential stress is then determined
	self-consistently by the field equations.

	These results show that specifying a dark matter density profile is not
	sufficient to fix strong-field orbital properties. Even with the same
	truncated Hernquist density profile, models characterized by different
	energy-momentum tensors can produce different and potentially opposite
	ISCO corrections.
	For the parameter ranges considered here, the ISCO shifts are small, reaching
	at most a few times $10^{-2}M$ in Table~\ref{tab:isco}. We therefore
	interpret them as theoretical differences in strong-field orbital properties
	between models characterized by different energy-momentum tensors, rather
	than predictions of directly detectable deviations.

\section{General First-Order ISCO Response}
\label{sec:isco_response}
\subsection{Density and radial-pressure coefficients}

For a static spherical geometry, the Einstein equations give
\begin{equation}
m'(r)=4\pi r^2\rho(r),\qquad
\frac{f'(r)}{f(r)}
=\frac{2[m(r)+4\pi r^3p_r(r)]}{r[r-2m(r)]}.
\label{eq:temporal-metric-pressure}
\end{equation}
At fixed central mass and density, the mass function and radial metric
are fixed, whereas the temporal metric also depends on the radial pressure.
To quantify this dependence, we retain the Bardeen source with
$p_r^{\rm Bar}=-\rho_{\rm Bar}$ and introduce the constant effective closure
\begin{equation}
p_r^{\rm DM}=\lambda\rho_{\rm DM}.
\label{eq:lambda-closure}
\end{equation}
Models~2 and~3 correspond to $\lambda=-1$ and $\lambda=0$, respectively.
The interval $-1\leq\lambda\leq0$ connects these two prescriptions; the
response formula below also applies to other fixed finite values of
$\lambda$ when the perturbative assumptions hold. Intermediate values
describe an effective anisotropic fluid, not necessarily a collisionless
particle model.

Define $x=r/M$ and $\hat g_s=g_s/M$, and write the isolated Bardeen
background as
\begin{align}
F_{\hat g_s}(x)&=1-\frac{2x^2}{(x^2+\hat g_s^2)^{3/2}},\nonumber\\
U(x)&=\ln F_{\hat g_s}(x),\qquad D(x)=xF_{\hat g_s}(x).
\label{eq:bardeen-dimensionless-functions}
\end{align}
The background ISCO $x_I(\hat g_s)$ is the physical root of
\begin{equation}
\mathcal H[U]\equiv xU''+3U'-x(U')^2=0,
\label{eq:bardeen-isco-equation}
\end{equation}
and the cutoff $x_t=r_t/M$ is determined by the marginally bound condition
\begin{equation}
\frac{2F_{\hat g_s}(x_t)^2}
{2F_{\hat g_s}(x_t)-x_tF_{\hat g_s}'(x_t)}=1.
\label{eq:bardeen-cutoff-equation}
\end{equation}
All primes in the remainder of this subsection denote derivatives with
respect to $x$ at fixed $\hat g_s$ and fixed profile parameters.
We introduce a positive perturbation amplitude $\epsilon$ through
\begin{align}
\frac{m_{\rm DM}(r)}M&=\epsilon\mu(x),\nonumber\\
\mathcal R(x)&=\frac{4\pi M^2}{\epsilon}\rho_{\rm DM}(Mx),\nonumber\\
\mu'(x)&=x^2\mathcal R(x).
\label{eq:general-halo-mass-perturbation}
\end{align}
The complete truncated density is the input, with
$\mu(x)=\int_{x_t}^x y^2\mathcal R(y)\,dy$ outside the cutoff and zero halo
mass inside. Its dependence on the magnetic charge through $x_t$ is
retained. No additional cutoff correction is introduced.

Writing $\ln f_\lambda=U+\epsilon h_\lambda+O(\epsilon^2)$ locally,
up to a constant normalization irrelevant to the ISCO, expansion of
Eq.~\eqref{eq:temporal-metric-pressure} gives
\begin{align}
K_\lambda\equiv h_\lambda'
&=K_\rho+\lambda K_p,\nonumber\\
K_\rho&=\frac{2\mu D'}{D^2},\qquad
K_p=\frac{2\mu'}D=\frac{2x^2\mathcal R}D.
\label{eq:general-pressure-kernel}
\end{align}
Here $K_p$ denotes the pressure response per unit $\lambda$.
Thus the closure parameter remains explicit throughout the perturbation
calculation. The perturbations and their relevant derivatives must be
small near $x_I$; $\mu$ is assumed twice differentiable there.
The central parameters $M$ and $g_s$ are held fixed.

Let $A=3-2xU'$ and
\begin{equation}
Q=\partial_x\mathcal H[U]
=xU'''+4U''-(U')^2-2xU'U''.
\label{eq:bardeen-isco-denominator}
\end{equation}
For a nondegenerate background ISCO, $Q(x_I)\neq0$, linearizing
Eq.~\eqref{eq:bardeen-isco-equation} gives
\begin{align}
\frac{r_{\rm ISCO}(\lambda)}M
&=x_I(\hat g_s)+\epsilon C_\lambda+O(\epsilon^2),
\label{eq:bardeen-isco-expansion}\\
C_\lambda[\mu;\hat g_s]
&=-\left.\frac{xK_\lambda'+AK_\lambda}{Q}\right|_{x_I}
=C_\rho+\lambda C_p,
\label{eq:general-density-pressure-criterion}
\end{align}
where the explicit density and pressure response coefficients are
\begin{align}
C_\rho&=-\left.\frac{2}{Q}
\left[\frac{x\mu'D'+x\mu D''+A\mu D'}{D^2}
-\frac{2x\mu(D')^2}{D^3}\right]\right|_{x_I},\nonumber\\
C_p&=-\left.\frac{2}{Q}
\left[\frac{x\mu''+A\mu'}D-\frac{x\mu'D'}{D^2}\right]\right|_{x_I}.
\label{eq:bardeen-response-coefficients}
\end{align}
Here, $C_\rho$ is the contribution of the dark matter density to the
ISCO shift, while $\lambda C_p$ is the contribution of the dark matter
radial pressure.
Consequently, the two models have
\begin{equation}
C_2=C_{\lambda=-1}=C_\rho-C_p,\qquad
C_3=C_{\lambda=0}=C_\rho.
\end{equation}
For $\epsilon>0$, $C_\lambda>0$ predicts an outward shift and
$C_\lambda<0$ an inward shift relative to the isolated Bardeen ISCO at
the same charge. At $C_\lambda=0$, only the first-order displacement
vanishes. Positivity of the density alone therefore does not fix the sign.

In the Schwarzschild limit, $x_I=6$ and $Q(x_I)=1/48$, giving
\begin{equation}
C_\lambda[\mu;0]
=\left[6\mu-36\mu'
-\lambda(12\mu'+144\mu'')\right]_{x=6}.
\label{eq:schwarzschild-general-density-response}
\end{equation}
For the truncated Hernquist profile with $a_0\gg M$, choosing
$\epsilon=MM_{\rm DM}/a_0^2$ yields
\begin{equation}
\begin{aligned}
\frac{m(r)-m_{\rm Bar}(r)}M
&=\frac{a_0M_{\rm DM}M(x-x_t)^2}
{(a_0+Mx)^2(a_0+Mx_t)}
\\
&=\epsilon(x-x_t)^2
+O\!\left(\epsilon\frac{M}{a_0}\right).
\end{aligned}
\label{eq:bardeen-mass-dilute-limit}
\end{equation}
Hence $\mu_{\rm H}=(x-x_t)^2$ at leading order, and
\begin{equation}
K_\lambda
=\frac{2(x-x_t)^2D'}{D^2}
+\frac{4\lambda(x-x_t)}D.
\label{eq:hernquist-lambda-kernel}
\end{equation}
The remainder in Eq.~\eqref{eq:bardeen-isco-expansion} then also includes
$O(\epsilon M/a_0)$. At $\hat g_s=0$, $x_t=4$ and
$C_\lambda=-120-336\lambda$, reproducing $C_2=216$ and $C_3=-120$.
The latter agrees with the Einstein cluster result for $r_t=4M$
in Ref.~\cite{Shen2025}. At nonzero charge, solving for $x_I$ and $x_t$
and substituting into the same response formula gives
Table~\ref{tab:bardeen-response}. For example, at $\hat g_s=0.5$ and
$\epsilon=10^{-4}$, the first-order predictions are
$r_{\rm ISCO}^{(2)}/M=5.59015$ and
$r_{\rm ISCO}^{(3)}/M=5.56164$, consistent with the small higher-order
differences in Table~\ref{tab:isco}.

\begin{table}[tb]
\centering
\caption{Leading response coefficients for the truncated Hernquist
profile. Models~2 and~3 correspond to $\lambda=-1$ and 0.
The last column gives the critical value in
Eq.~\eqref{eq:lambda-critical}.}
\label{tab:bardeen-response}
\begin{ruledtabular}
\begin{tabular}{cccccc}
$\hat g_s$ & $x_I$ & $x_t$ & $C_2$ & $C_3$ & $\lambda_c$\\
0.0 & 6.00000 & 4.00000 & 216.000 & -120.000 & $-0.35714$\\
0.3 & 5.85376 & 3.89578 & 202.858 & -114.903 & $-0.36160$\\
0.5 & 5.57222 & 3.69306 & 179.274 & -105.879 & $-0.37131$
\end{tabular}
\end{ruledtabular}
\end{table}

\subsection{Critical radial-pressure closure}

The linear response also determines the critical closure directly:
\begin{equation}
\lambda_c[\mu;\hat g_s]=-\frac{C_\rho}{C_p}
=\frac{C_3}{C_2-C_3},\qquad C_p\neq0.
\label{eq:lambda-critical}
\end{equation}
For distinct endpoint coefficients, $\lambda_c$ lies in $[-1,0]$
precisely when $C_2C_3\leq0$. If $C_p=0$, there is no unique critical
value: the first-order shift is independent of $\lambda$.
For the Hernquist profile, $\lambda_c$ becomes slightly more negative
at the charges listed in Table~\ref{tab:bardeen-response}.
With a common asymptotic normalization, the radial Einstein equation
also gives the exact metric relation
\begin{equation}
f_\lambda=f_3^{1+\lambda}f_2^{-\lambda},
\label{eq:lambda-metric-interpolation}
\end{equation}
while the first-order ISCO coefficient satisfies
\begin{equation}
C_\lambda=(1+\lambda)C_3-\lambda C_2.
\label{eq:lambda-isco-interpolation}
\end{equation}
The vanishing first-order ISCO shift at $\lambda_c$ does not imply an
unchanged temporal metric or orbital frequency.

\subsection{Validation across density profiles and backgrounds}

Equation~\eqref{eq:general-density-pressure-criterion} provides a direct
procedure for determining the leading ISCO response to a prescribed
spherical density and radial pressure at fixed magnetic charge. For
$p_r^{\rm DM}=\lambda\rho_{\rm DM}$ with constant $\lambda$, one inserts
the complete density profile, including any cutoff, into $\mu(x)$ and
evaluates $C_\lambda$ at the isolated Bardeen ISCO. For a general prescribed
pressure, the same linear response operator applies with
$K=2\mu D'/D^2+2x^2\mathcal P/D$, where
$\mathcal P=4\pi M^2p_r^{\rm DM}/\epsilon$.
A positive coefficient predicts an outward shift and a negative coefficient
an inward shift, without first solving for the full perturbed temporal
metric. This procedure applies whenever the local perturbations are small,
the required derivatives exist, and the background ISCO is nondegenerate;
a vanishing first-order coefficient requires higher-order analysis.

We verified the criterion against the standard unexpanded ISCO calculation,
in which the complete mass and pressure are inserted into
$W=d\ln f/dx$ from Eq.~\eqref{eq:temporal-metric-pressure} and the full
condition $xW'+3W-xW^2=0$ is solved numerically. We considered the
untruncated and linearly truncated Navarro--Frenk--White, Hernquist,
Plummer, cored Dehnen, and Jaffe profiles at the three representative
magnetic charges, for both $\lambda=-1$ and $0$. In all 60 cases the
first-order criterion predicts the same shift direction as the unexpanded
calculation. Excluding the strongly cusped Jaffe profile, the largest
relative difference in the shift magnitude is approximately
$2.2\times10^{-2}\%$.

Selected signed shifts are listed in
Table~\ref{tab:profile-isco-validation}. The first-order results are
obtained from Eq.~\eqref{eq:general-density-pressure-criterion}, while
the full results follow from the unexpanded ISCO equation. The table
uses the Navarro--Frenk--White profile to illustrate close agreement and
lists explicitly the less accurate Jaffe cases discussed below. Positive
and negative entries denote outward and inward shifts, respectively.
\begin{table*}[t]
\caption{Selected comparisons of the first-order analytic and full
unexpanded ISCO shifts. Except for the last row, $M_{\rm DM}/M=10^2$ and
$a_0/M=10^4$. The shifts are measured relative to the isolated ISCO of
the indicated background and are given in units of $M$. Here $q$ denotes
$\hat g_s$ for Bardeen and $\hat\ell$ for Hayward, ``Untr.'' denotes the
untruncated parent profile, and $n=1$ denotes a linear cutoff. The last
row uses $\rho_0M^2=0.01$ and $r_s/M=0.2$ from
Ref.~\cite{Rani2025Dehnen}. The shift directions agree in every row; the
last column is defined by
$\delta_{\rm shift}=|\Delta x_{\rm ana}-\Delta x_{\rm full}|/
|\Delta x_{\rm full}|\times100\%$. It therefore measures the accuracy
of the first-order halo-induced ISCO shift, not the relative difference
between the total ISCO radii.}
\label{tab:profile-isco-validation}
\begin{ruledtabular}
\begin{tabular}{cccccccc}
Background & Profile & Cutoff & $q$ & $\lambda$
& $\Delta x_{\rm ana}$ & $\Delta x_{\rm full}$
& Relative shift error \\
\hline
Bardeen & Navarro--Frenk--White & Untr. & 0.5 & $-1$
& $ 6.03598\times10^{-5}$ & $ 6.03628\times10^{-5}$ & $0.00493\%$ \\
Bardeen & Navarro--Frenk--White & Untr. & 0.5 & $0$
& $-5.71059\times10^{-5}$ & $-5.71022\times10^{-5}$ & $0.00663\%$ \\
Bardeen & Navarro--Frenk--White & $n=1$ & 0.5 & $-1$
& $ 6.01095\times10^{-5}$ & $ 6.01122\times10^{-5}$ & $0.00446\%$ \\
Bardeen & Navarro--Frenk--White & $n=1$ & 0.5 & $0$
& $-3.55178\times10^{-5}$ & $-3.55167\times10^{-5}$ & $0.00308\%$ \\
\hline
Bardeen & Jaffe & Untr. & 0.0 & $0$
& $ 2.15741\times10^{-4}$ & $ 1.95774\times10^{-2}$ & $98.9\%$ \\
Bardeen & Jaffe & Untr. & 0.3 & $0$
& $ 1.41601\times10^{-2}$ & $ 3.19419\times10^{-2}$ & $55.7\%$ \\
Bardeen & Jaffe & Untr. & 0.5 & $0$
& $ 4.18368\times10^{-2}$ & $ 5.66484\times10^{-2}$ & $26.1\%$ \\
\hline
Hayward & Hernquist & $n=1$ & 0.30 & $-1$
& $ 2.11450\times10^{-4}$ & $ 2.11481\times10^{-4}$ & $0.0149\%$ \\
Hayward & Hernquist & $n=1$ & 0.30 & $0$
& $-1.18903\times10^{-4}$ & $-1.18891\times10^{-4}$ & $0.0100\%$ \\
\hline
Schwarzschild & Dehnen~\cite{Rani2025Dehnen} & Untr. & -- & $-1$
& $ 2.53047\times10^{-1}$ & $ 2.63660\times10^{-1}$ & $4.03\%$
\end{tabular}
\end{ruledtabular}
\end{table*}

The Jaffe cases illustrate the expected limitation of the expansion. For
the linearly truncated profile, the first-order and full shift magnitudes differ by approximately
$4$--$6\%$, while the direction remains correct. For the untruncated
Einstein cluster case, the leading density response nearly cancels; the
relative differences at $\hat g_s=0$, $0.3$, and $0.5$ are approximately
$99\%$, $56\%$, and $26\%$, respectively, even though both methods still
predict an outward shift. Thus the first-order formula accurately
determines both sign and magnitude for dilute profiles away from a leading-order
cancellation; close to $C_\lambda=0$, the full ISCO equation is required.

The response formula is not tied to the Bardeen background. Replacing
$F_{\hat g_s}$ in Eqs.~\eqref{eq:bardeen-dimensionless-functions}--
\eqref{eq:general-density-pressure-criterion} by the Hayward lapse
$F_{\rm H}=1-2x^2/(x^3+2\hat\ell^2)$ gives the Hayward results listed in
Table~\ref{tab:profile-isco-validation}. As an independent comparison
with a published value, the Dehnen-halo metric of
Ref.~\cite{Rani2025Dehnen}, which has $p_r=-\rho$, is obtained by choosing
$\epsilon=32\pi(\rho_0M^2)(r_s/M)^2$ and
$\mu=x\sqrt{1+(r_s/M)/x}/2$. The corresponding first-order and published
full results are also compared in Table~\ref{tab:profile-isco-validation}.
Together, these checks show that the first-order criterion applies
to small static, spherically symmetric matter perturbations of different
nondegenerate black-hole backgrounds, rather than only to the Bardeen--
Hernquist system.

\section{Effects of magnetic charge and dark matter pressure on EMRI dephasing}
\label{sec:magnetic_charge_dephasing}
\subsection{Evolution framework and common initial conditions}

We next investigate how the Bardeen magnetic charge and the dark matter radial-pressure prescription modify the accumulated phase of a quasicircular EMRI through their effects on the background spacetime geometry. We follow the conservative sector of the fully relativistic framework for EMRIs in dark matter environments developed by Fier et al.~\cite{Fier:2026yjc}, evaluating the circular-orbit quantities in each background metric and modeling the gravitational-wave energy loss at leading quadrupolar order. To isolate the effects of the magnetic charge and radial-pressure prescription, we keep the dark matter halo parameters fixed throughout this subsection as
\begin{equation}
	M_{\rm DM}=10^{12}M_{\odot},
	\qquad
	a_{0}=20~{\rm kpc}.
	\label{eq:fixed_halo_charge_parameters}
\end{equation}
These values represent a massive Milky-Way-like galactic halo rather
than a precise model of the Milky Way. The binary masses
and observation time are chosen as
\begin{equation}
	M=10^{6}M_{\odot},
	\qquad
	m_{\star}=10M_{\odot},
	\qquad
	T_{\rm obs}=4~{\rm yr}.
	\label{eq:charge_emri_parameters}
\end{equation}
The primary mass is a representative benchmark for an EMRI observable
by the Laser Interferometer Space Antenna (LISA), rather than the
measured mass of Sagittarius~A*. We consider the dimensionless
magnetic charges
\begin{equation}
	\frac{g_s}{M}
	=
	0,\;0.01,\;0.05,\;0.1,\;0.3,\;0.5.
	\label{eq:magnetic_charge_values}
\end{equation}
Dynamical friction and accretion are neglected in this calculation,
so that the results isolate the conservative modification of the
orbital dynamics produced by the magnetic charge and the
background geometry modified by dark matter.

For a static and spherically symmetric spacetime,
\begin{equation}
	ds^{2}
	=
	-f(r)dt^{2}
	+\frac{dr^{2}}{g(r)}
	+r^{2}d\Omega^{2},
\end{equation}
the angular frequency and specific energy of a circular timelike
geodesic are
\begin{equation}
	\Omega^{2}(r)
	=
	\frac{f'(r)}{2r},
	\label{eq:charge_orbital_frequency}
\end{equation}
and
\begin{equation}
	{\cal E}^{2}(r)
	=
	\frac{2f^{2}(r)}
	{2f(r)-rf'(r)},
	\label{eq:charge_specific_energy}
\end{equation}
respectively. At leading quadrupolar order, the gravitational-wave
luminosity is approximated by
\begin{equation}
	{\cal F}_{\rm GW}
	=
	\frac{32}{5}
	m_{\star}^{2}r^{4}\Omega^{6}.
	\label{eq:charge_gw_flux}
\end{equation}
The adiabatic orbital evolution follows from energy balance,
\begin{equation}
	\dot r
	=
	-\frac{{\cal F}_{\rm GW}}
	{m_{\star}\,d{\cal E}/dr},
	\label{eq:charge_radial_evolution}
\end{equation}
and the accumulated phase of the dominant quadrupolar harmonic is
\begin{equation}
	\Phi_{\rm GW}
	=
	2\int_{0}^{T_{\rm obs}}
	\Omega[r(t)]\,dt.
	\label{eq:charge_accumulated_phase}
\end{equation}
Since the orbital radius decreases monotonically, the observation
time and accumulated phase may equivalently be written as
\begin{align}
	T_{\rm obs}
	&=
	\int_{r_f}^{r_i}
	\frac{m_{\star}\,d{\cal E}/dr}
	{{\cal F}_{\rm GW}}\,dr,
	\label{eq:charge_time_integral}
	\\
	\Phi_{\rm GW}
	&=
	2\int_{r_f}^{r_i}
	\Omega(r)
	\frac{m_{\star}\,d{\cal E}/dr}
	{{\cal F}_{\rm GW}}\,dr.
	\label{eq:charge_phase_integral}
\end{align}
These equations implement the conservative part of the framework of
Fier et al.~\cite{Fier:2026yjc}. In contrast to the additional scalar
radiation considered in the original accumulated dephasing application
of Maselli et al.~\cite{Maselli:2020scalar}, the phase differences here
arise from changes in the background geometry within the adopted
quadrupole-flux approximation.

\subsubsection{Common initial frequency and ISCO regulator}
\label{sec:charge_initial_frequency}

Following Ref.~\cite{Fier:2026yjc}, we choose the initial orbit such
that the uncharged vacuum inspiral reaches the vicinity of the
Schwarzschild ISCO after four years. Numerically, the reference
evolution is terminated at
\begin{equation}
	r_{\rm end}
	=
	6M+10^{-6}M,
	\label{eq:charge_isco_regulator}
\end{equation}
to avoid evaluating Eq.~(\ref{eq:charge_radial_evolution}) exactly at
$d{\cal E}/dr=0$ and to keep the endpoint on the stable circular-orbit
branch. This is a numerical regulator, not a physical cutoff; changing
it from $10^{-3}M$ to $10^{-6}M$ alters the largest dephasing by only
$1.7\times10^{-2}$ rad.

This reference evolution determines the common initial gravitational-wave
frequency
\begin{equation}
	f_{{\rm GW},i}
	=\frac{\Omega_i}{\pi}=
	1.53166191~{\rm mHz}.
	\label{eq:common_initial_charge_frequency}
\end{equation}
Because the metric changes the frequency--radius relation, the initial
radius of each configuration is determined separately from
\begin{equation}
	\Omega_{g_s}(r_{i,g_s})=\Omega_i.
	\label{eq:same_initial_charge_frequency}
\end{equation}
Thus all inspirals start at the same observed frequency and phase and
are evolved for the same duration.

\subsection{Dephasing caused by the magnetic charge}

The marginally bound cutoff radius and the ISCO are also recalculated
for every value of the magnetic charge. The phase difference caused by
the charge is defined separately within each dark matter model as
\begin{equation}
	\Delta\Phi_{g_s}^{(k)}
	\equiv
	\Phi_k(g_s)-\Phi_k(0),
	\label{eq:charge_phase_difference}
\end{equation}
where $k=1,2,3$ labels the halo model. The $g_s=0$ configuration of
each model is therefore its own reference and satisfies
$\Delta\Phi_{g_s}^{(k)}=0$ by definition.

\begin{table*}[t]
	\caption{
		Orbital parameters and accumulated phase differences induced by the charge
		for fixed dark matter parameters
		$M_{\rm DM}=10^{12}M_{\odot}$ and $a_0=20~{\rm kpc}$.
		All configurations begin at the same gravitational-wave frequency,
		$f_{{\rm GW},i}=1.53166191~{\rm mHz}$, and are evolved for four
		years. Models~1 and 2 give identical results at the displayed
		precision and are therefore combined. The phase difference is
		defined relative to the uncharged configuration of the same
		dark matter model.
	}
	\label{tab:fixed_halo_charge_dephasing}
	\begin{ruledtabular}
		\begin{tabular}{ccccccc}
			Model
			&
			$g_s/M$
			&
			$r_{\rm cut}/M$
			&
			$r_{\rm ISCO}/M$
			&
			$r_i/M$
			&
			$r_f/M$
			&
			$\Delta\Phi_{g_s}~[{\rm rad}]$
			\\
			\hline
			1 and 2
			& 0
			& 4.000000
			& 6.000000
			& 12.119454
			& 6.000001
			& 0
			\\
			1 and 2
			& 0.01
			& 3.999887
			& 5.999842
			& 12.119441
			& 6.044850
			& $-47.6967$
			\\
			1 and 2
			& 0.05
			& 3.997185
			& 5.996039
			& 12.119144
			& 6.220000
			& $-1115.9082$
			\\
			1 and 2
			& 0.10
			& 3.988715
			& 5.984123
			& 12.118216
			& 6.429593
			& $-4121.3344$
			\\
			1 and 2
			& 0.30
			& 3.895785
			& 5.853764
			& 12.108298
			& 7.172859
			& $-27765.7894$
			\\
			1 and 2
			& 0.50
			& 3.693058
			& 5.572225
			& 12.088383
			& 7.786197
			& $-59943.0058$
			\\
			\hline
			3
			& 0
			& 4.000000
			& 6.000000
			& 12.119434
			& 6.015739
			& 0
			\\
			3
			& 0.01
			& 3.999887
			& 5.999842
			& 12.119422
			& 6.047514
			& $-47.5784$
			\\
			3
			& 0.05
			& 3.997185
			& 5.996039
			& 12.119125
			& 6.220541
			& $-1115.2393$
			\\
			3
			& 0.10
			& 3.988715
			& 5.984123
			& 12.118196
			& 6.429860
			& $-4120.0560$
			\\
			3
			& 0.30
			& 3.895785
			& 5.853764
			& 12.108279
			& 7.172940
			& $-27762.7357$
			\\
			3
			& 0.50
			& 3.693058
			& 5.572225
			& 12.088364
			& 7.786240
			& $-59938.8374$
			\\
		\end{tabular}
	\end{ruledtabular}
\end{table*}

Table~\ref{tab:fixed_halo_charge_dephasing} shows that both the
marginally bound cutoff radius and the ISCO decrease continuously as
the magnetic charge increases, while every trajectory remains outside
the ISCO. The charge changes the frequency--radius relation, circular-orbit
energy, and luminosity; within the adopted balance law, the charged
systems traverse less of the high-frequency strong-field region and
therefore accumulate fewer cycles, yielding
$\Delta\Phi_{g_s}^{(k)}<0$.

For small magnetic charge, the numerical results are accurately
described by
\begin{equation}
	\Delta\Phi_{g_s}^{(k)}
	\simeq
	c_2^{(k)}
	\left(\frac{g_s}{M}\right)^2.
	\label{eq:small_charge_quadratic_scaling}
\end{equation}
This follows from the even dependence of the Bardeen metric on $g_s$;
higher-order effects become relevant beyond the small-charge regime.
The largest tabulated dephasing is about $3.7\%$ of the total four-year
phase, so its large absolute value primarily reflects coherent EMRI
phase accumulation.

Models~1 and 2 produce identical results at the displayed precision.
Both obey $p_r^{\rm DM}=-\rho_{\rm DM}$ and $f=g$, and the adopted
halo is extremely dilute in the sampled strong-field region; hence
the cutoff has a negligible effect on their accumulated phases.

\subsection{Dephasing caused by the radial-pressure prescription}
\label{sec:pressure_induced_dephasing}

The phase differences in
Eq.~(\ref{eq:charge_phase_difference}) are defined relative to the
uncharged configuration within each model and therefore primarily
measure the effect of the magnetic charge. To isolate the influence of the
dark matter radial pressure, we directly compare Models~2 and 3.
These models have the same truncated density profile, the same
dark matter mass function, and the same cutoff radius for each value of
the charge. Their radial-pressure prescriptions are, however,
\begin{equation}
	p_{r,2}^{\rm DM}=-\rho_{\rm DM},
	\qquad
	p_{r,3}^{\rm DM}=0.
	\label{eq:two_pressure_prescriptions}
\end{equation}
Their comparison therefore separates the effect of the radial pressure
from that of the density profile and inner cutoff.

For each fixed magnetic charge, we define the phase difference due to
the radial-pressure prescription as
\begin{equation}
	\Delta\Phi_{\rm pressure}(g_s)
	\equiv
	\Phi_3(g_s)-\Phi_2(g_s).
	\label{eq:pressure_phase_difference}
\end{equation}
The two inspirals compared in Eq.~(\ref{eq:pressure_phase_difference})
have the same binary masses, magnetic charge, initial
gravitational-wave frequency, initial phase, observation time,
dark matter density profile, and cutoff radius. Their initial
coordinate radii are determined separately using
Eq.~(\ref{eq:same_initial_charge_frequency}), because the two temporal
metric functions give slightly different relations between frequency and radius.

\begin{table}[t]
	\caption{
		EMRI phase difference due to pressure between Models~3 and 2.
		The two models have the same truncated density profile but satisfy
		$p_{r,3}^{\rm DM}=0$ and
		$p_{r,2}^{\rm DM}=-\rho_{\rm DM}$, respectively. The dark matter
		parameters are fixed at $M_{\rm DM}=10^{12}M_{\odot}$ and
		$a_0=20~{\rm kpc}$. All inspirals begin at the same gravitational-wave
		frequency and are evolved for four years.
	}
	\label{tab:pressure_phase_difference}
	\begin{ruledtabular}
		\begin{tabular}{ccc}
			$g_s/M$
			&
			$\Phi_3-\Phi_2~[{\rm rad}]$
			&
			$(\Phi_3-\Phi_2)/(2\pi)$
			\\
			\hline
			0
			& $-7.9251$
			& $-1.2613$
			\\
			0.01
			& $-7.8068$
			& $-1.2425$
			\\
			0.05
			& $-7.2562$
			& $-1.1549$
			\\
			0.10
			& $-6.6467$
			& $-1.0578$
			\\
			0.30
			& $-4.8714$
			& $-0.7753$
			\\
			0.50
			& $-3.7566$
			& $-0.5979$
			\\
		\end{tabular}
	\end{ruledtabular}
\end{table}

Table~\ref{tab:pressure_phase_difference} demonstrates that the choice
of radial pressure produces several radians of dephasing even at fixed
density and cutoff. The negative sign means that Model~3 accumulates
fewer gravitational-wave radians than Model~2, while the decreasing
absolute value with increasing $g_s$ shows that the magnetic charge
also modifies the response to the pressure prescription.

The origin of this difference follows from
\begin{equation}
	\frac{d}{dr}
	\ln\left(\frac{f}{g}\right)
	=
	\frac{8\pi r(\rho+p_r)}{g}.
	\label{eq:lapse_pressure_relation_emri}
\end{equation}
Model~2 has $\rho_{\rm DM}+p_{r,2}^{\rm DM}=0$ and hence $f_2=g_2$,
whereas Model~3 has $\rho_{\rm DM}+p_{r,3}^{\rm DM}=\rho_{\rm DM}$
and $f_3\ne g_3$. The resulting change in the temporal metric alters
$\Omega$, ${\cal E}$, the luminosity, and therefore the accumulated
phase despite the common density profile.

The phase difference due to pressure is related to the
dephasings caused by the charge in the individual models by
\begin{align}
	\Delta\Phi_{g_s}^{(3)}
	-
	\Delta\Phi_{g_s}^{(2)}
	&=
	\Delta\Phi_{\rm pressure}(g_s)
	\nonumber\\
	&\quad -
	\Delta\Phi_{\rm pressure}(0).
	\label{eq:charge_pressure_relation}
\end{align}
Thus the difference between the two charge-induced dephasings measures
only the change in the pressure effect relative to $g_s=0$, whereas
Table~\ref{tab:pressure_phase_difference} gives the full pressure-induced
phase difference at fixed charge. Although subdominant to the magnetic-charge
effect, it accumulates to several radians and shows that the density
profile alone does not determine the gravitational-wave phase.

These are fixed-parameter results within the leading-order adiabatic
quadrupole-flux approximation, not a detector-level measurability
forecast. Such a forecast requires relativistic fluxes, an
inspiral-to-plunge transition, waveform overlaps, and simultaneous
estimation of source and halo parameters.

\section{Summary}
\label{sec:summary}

We compared three static spherical models of a Bardeen black hole in a
Hernquist halo: an untruncated halo with $p_r^{\rm DM}=-\rho_{\rm DM}$,
its truncated counterpart, and a truncated Einstein cluster with
$p_r^{\rm DM}=0$. The two truncated models share the same density,
mass function, and cutoff, fixed by the isolated marginally bound orbit.
Their comparison isolates the radial-pressure contribution to the
temporal metric and orbital dynamics. In the examples studied, the
Einstein cluster has positive tangential pressure in the displayed
exterior region and violates the dominant energy condition only near
the cutoff, whereas the other closures have different stress and
energy-condition behavior. Increasing the magnetic charge moves the
cutoff, photon sphere, and ISCO inward over the parameter range examined.

The first-order criterion $C_\lambda=C_\rho+\lambda C_p$ separates the
density and radial-pressure contributions to the halo-induced ISCO
shift. For the truncated Hernquist profile, it explains the outward
shift in Model~2 and inward shift in Model~3. Tests of five density
profiles, with and without a linear cutoff, recover the full numerical
shift directions; the Jaffe cases show the loss of relative accuracy
when the leading response nearly cancels. The Hayward test and the
published Dehnen-halo comparison further support applicability beyond
the Bardeen--Hernquist system within the stated perturbative assumptions.
For the published example, the discrepancy is $4.03\%$ of the halo-induced
shift, equivalent to $0.17\%$ of the total ISCO radius.
The critical closure $\lambda_c=-C_\rho/C_p$ identifies a vanishing
leading shift, but does not eliminate changes in the temporal metric
or orbital frequency.

For four-year quasicircular EMRIs with $M=10^6M_\odot$,
$m_\star=10M_\odot$, $M_{\rm DM}=10^{12}M_\odot$, and
$a_0=20~{\rm kpc}$, the magnetic-charge-induced dephasing ranges from
approximately $-47.7$ rad at $g_s/M=0.01$ to $-5.99\times10^4$ rad at
$g_s/M=0.5$, with approximately quadratic behavior for $g_s/M\leq0.1$.
Models~1 and 2 agree at the displayed precision for this dilute halo.
At fixed charge, changing the radial pressure from $-\rho_{\rm DM}$ to
zero gives $\Phi_3-\Phi_2\simeq-7.93$ rad at $g_s=0$ and $-3.76$ rad at
$g_s/M=0.5$. Thus the pressure prescription affects the accumulated phase
even at fixed density and cutoff, and the magnitude of this effect
decreases with charge. The difference between the models'
charge-induced dephasings remains below approximately $4.2$ rad and
measures the change in this pressure effect relative to $g_s=0$.

The response criterion requires small local perturbations, sufficient
smoothness, and a nondegenerate background ISCO; it does not imply a
universal shift direction. The EMRI results are fixed-parameter estimates
within a leading-order adiabatic quadrupole-flux approximation.
Assessing observational distinguishability requires relativistic
fluxes, a transition-to-plunge treatment, and waveform comparisons
accounting for source and halo parameter degeneracies. Further
extensions include dynamically determined inner depletion, a smooth
cutoff, and rotating or time-dependent halos.

\begin{acknowledgments}
This work was supported by the National Natural Science Foundation of
China under Grants No.~12205129 and No.~12247101; the Fundamental
Research Funds for the Central Universities under Grants
No.~lzujbky-2025-it05 and No.~lzujbky-2025-jdzx07; the Natural Science
Foundation of Gansu Province under Grants No.~22JR5RA389 and
No.~25JRRA799; the 111 Center under Grant No.~B20063; and the Key
Project of the Department of Education of Hunan Province under Grant
No.~25A0084. Wen-Di Guo was supported by the Talent Scientific Fund of
Lanzhou University.
\end{acknowledgments}

	\appendix
	\label{app:energy}
\section{Analytic Expressions for the Energy Conditions of Models 1 and 2}

For an anisotropic fluid, the energy conditions considered in this work are
\begin{align}
	&\text{WEC:}\quad \rho\ge 0,\qquad \rho+p_r\ge 0,\qquad \rho+p_t\ge 0, \label{A1}\\
	&\text{NEC:}\quad \rho+p_r\ge 0,\qquad \rho+p_t\ge 0, \label{A2}\\
	&\text{SEC:}\quad \rho+p_r\ge 0,\qquad \rho+p_t\ge 0,\qquad \rho+p_r+2p_t\ge 0, \label{A3}\\
	&\text{DEC:}\quad \rho\ge 0,\qquad |p_r|\le \rho,\qquad |p_t|\le \rho. \label{A4}
\end{align}

For both Models 1 and 2 the closure $p_r=-\rho$ holds for the total energy-momentum
tensor, so that $g_{tt}=-1/g_{rr}$ and
\begin{equation}
	\rho+p_r\equiv 0. \label{A5}
\end{equation}
From the conservation of the energy-momentum tensor,
$p_r'=-(\rho+p_r)\Phi' + \dfrac{2}{r}(p_t-p_r)$ with $\Phi'=f'/2f$,
the tangential pressure is determined by the density profile alone:
\begin{equation}
	\boxed{\;p_t=-\rho-\frac{r}{2}\,\rho'\,.\;}
	\label{A6}
\end{equation}
Consequently, using (\ref{A5}) and (\ref{A6}),
\begin{equation}
	\rho+p_r+2p_t=2p_t,\qquad
	\rho+p_t=-\frac{r}{2}\rho',\qquad
	\rho-p_t=2\rho+\frac{r}{2}\rho'. \label{A7}
\end{equation}
The conditions (\ref{A3})--(\ref{A4}) therefore reduce to
\begin{equation}
	\rho\ge 0,\qquad 2p_t\ge 0,\qquad \rho-p_t\ge 0,\qquad \rho+p_t\ge 0, \label{A8}
\end{equation}
and we list the explicit inequalities below. Each inequality is written after
multiplying by the manifestly positive denominator, so the direction of the
inequality is preserved.

\subsection*{1. Model 1}

The total density is
\begin{equation}
	\rho_1(r)=\frac{3Mg_s^2}{4\pi(r^2+g_s^2)^{5/2}}
	+\frac{M_{\rm DM}a_0}{2\pi r(a_0+r)^3},
	\label{A9}
\end{equation}
and the tangential pressure follows from (\ref{A6}):
\begin{equation}
	p_{t,1}(r)=\frac{3Mg_s^2(3r^2-2g_s^2)}{8\pi(r^2+g_s^2)^{7/2}}
	+\frac{M_{\rm DM}a_0(2r-a_0)}{4\pi r(a_0+r)^4}.
	\label{A10}
\end{equation}

The strong energy condition $2p_t\ge 0$ is equivalent to
\begin{equation}
	3Mg_s^2\,r\,(3r^2-2g_s^2)(a_0+r)^4
	+2M_{\rm DM}a_0\,(2r-a_0)\,(r^2+g_s^2)^{7/2}\ge 0.
	\label{A11}
\end{equation}

For the dominant energy condition, the constraint $\rho-p_t\ge 0$ is equivalent to
\begin{equation}
	3Mg_s^2\,r\,(4g_s^2-r^2)(a_0+r)^4
	+6M_{\rm DM}a_0^2\,(r^2+g_s^2)^{7/2}\ge 0,
	\label{A12}
\end{equation}
and the constraint $\rho+p_t\ge 0$ is equivalent to
\begin{equation}
	15Mg_s^2\,r^3\,(a_0+r)^4
	+2M_{\rm DM}a_0\,(a_0+4r)\,(r^2+g_s^2)^{7/2}\ge 0.
	\label{A13}
\end{equation}

\subsection*{2. Model 2}

Outside the cutoff radius ($r\ge r_t$) the total density is
\begin{equation}
	\rho_2(r)=\frac{3Mg_s^2}{4\pi(r^2+g_s^2)^{5/2}}
	+\frac{M_{\rm DM}a_0}{2\pi r(a_0+r)^3}\left(1-\frac{r_t}{r}\right),
	\label{A14}
\end{equation}
and (\ref{A6}) gives
\begin{equation}
	p_{t,2}(r)=\frac{3Mg_s^2(3r^2-2g_s^2)}{8\pi(r^2+g_s^2)^{7/2}}
	+\frac{M_{\rm DM}a_0(2r-a_0-3r_t)}{4\pi r(a_0+r)^4}.
	\label{A15}
\end{equation}

The strong energy condition $2p_t\ge 0$ is equivalent to
\begin{equation}
	\frac{
		\begin{aligned}
			&3Mg_s^2\,r\,(a_0+r)^4\,(3r^2-2g_s^2)\\
			&\quad +2M_{\rm DM}a_0\,(r^2+g_s^2)^{7/2}\\
			&\qquad\times(2r-a_0-3r_t)
		\end{aligned}
	}{4\pi\,r\,(a_0+r)^4\,(r^2+g_s^2)^{7/2}}
	\ge 0.
	\label{A16}
\end{equation}

For the dominant energy condition, $\rho-p_t\ge 0$ is equivalent to
\begin{equation}
	\frac{
		\begin{aligned}
			&3Mg_s^2\,r^2\,(a_0+r)^4\,(4g_s^2-r^2)\\
			&\quad +2M_{\rm DM}a_0\,(r^2+g_s^2)^{7/2}\\
			&\qquad\times(3a_0r-2a_0r_t+r\,r_t)
		\end{aligned}
	}{8\pi\,r^2\,(a_0+r)^4\,(r^2+g_s^2)^{7/2}}
	\ge 0,
	\label{A17}
\end{equation}
and $\rho+p_t\ge 0$ is equivalent to
\begin{equation}
	\frac{
		\begin{aligned}
			&15Mg_s^2\,r^4\,(a_0+r)^4\\
			&\quad +2M_{\rm DM}a_0\,(r^2+g_s^2)^{7/2}\\
			&\qquad\times(a_0r+4r^2-2a_0r_t-5r\,r_t)
		\end{aligned}
	}{8\pi\,r^2\,(a_0+r)^4\,(r^2+g_s^2)^{7/2}}
	\ge 0.
	\label{A18}
\end{equation}

In the limit $r_t\to 0$, Eqs. (\ref{A16})--(\ref{A18}) reduce respectively to
(\ref{A11})--(\ref{A13}), as required by consistency with Model 1.
	
	\bibliography{ref}
	
\end{document}